\documentstyle[12pt]{article}
\newcommand{\bea}{\begin{eqnarray}}
\newcommand{\be}{\begin{equation}}
\newcommand{\eea}{\end{eqnarray}}
\newcommand{\ee}{\end{equation}}

\def\l{\lambda}
\def\m{\mu}

\def\G{\Gamma}
\def\J{\Psi}

\def\cc{{\cal C}}

\def\cm{{\cal M}}

\def\cq{{\cal Q}}
\def\car{{\cal R}}

\def\cz{{\cal Z}}

\begin{document}
\pagestyle{plain} \large \noindent . \vspace*{5mm} \\
CGPG-95/4-2\\gr-qc/9504014 \\April 1995
\begin{center} \LARGE \bf \vspace*{5mm}
A Modular Invariant Quantum Theory From the Connection Formulation of
(2+1)-Gravity on the Torus\\
\vspace*{10mm} \large \bf
 Peter Peld\'{a}n
\footnote{Email address: peldan@phys.psu.edu}\\
\vspace*{5mm} \large
 Center for Gravitational Physics and Geometry\\
Physics Department\\
The Pennsylvania State University, University Park, PA 16802, USA\\
\end{center}
\begin{abstract}
By choosing an unconventional polarization of the connection phase space in
(2+1)-gravity on the torus, a modular invariant quantum theory is constructed.
Unitary equivalence to the ADM-quantization is shown.
\\ PACS numbers: 04.60.-m, 04.60.Ds, 04.60.Kz \end{abstract}
\normalsize
It has recently been shown \cite{pp} that there are severe obstructions to
constructing a quantum theory for (2+1)-dimensional gravity on a manifold $\car
\times T^2$ in the connection representation if we treat all diffeomorphisms
as gauge. That is, we require physical states in our Hilbert space to be
invariant -- or at most transform according to a one-dimensional unitary
representation of the diffeomorphism group -- under all diffeomorphisms. In
\cite{pp} only the space-like sector \cite{carlip1, ashtekar1, louko} was
considered, but
the result can easily be transferred to the time-like sector as well.

The purpose of this Letter is to show that this problem can be solved
straightforwardly by choosing a different polarization of the phase space
\cite{woodhouse}.
The polarization chosen here leads automatically to the Heisenberg picture of
the ADM-formulation \cite{ADM, carlip2}.

The reason why we get such severe problems in trying to construct a quantum
theory in the connection representation with all diffeomorphisms treated as
gauge, is that the group of large diffeomorphisms has a very bad behavior on
the reduced configuration space; the group of large diffeomorphisms on the
torus does not act properly discontinuous in the space of gauge equivalent
classes of flat $SO(1,2)$ connections on the torus. However, we know that the
action of this group is nicely behaved in the phase space -- except for a
region of measure zero -- and we should therefore be able to quantize this
formulation with the use of a different polarization.\\ \\
In the space-like sector of (2+1)-gravity on the torus, we have, after
solving all constraints and gauge fixing all local symmetries, the following
theory \cite{carlip1, ashtekar1, louko}: the reduced phase space is the
cotangent
bundle over the reduced configuration space $\cq =(\car ^2-\{0,0\})/\cz _2$
coordinatized by the
coordinates $\m $ and $\l $, where $\cz _2$ acts in
$\car ^2$ as $\{\m , \l \}\rightarrow \{-\m ,-\l \}$.
The canonically conjugate momenta are denoted $a$ and $b$ and we have the two
fundamental Poisson brackets: $\{ \m ,a\}=\frac{1}{2}$ and $\{ \l
,b\}=-\frac{1}{2}$. Note that the $\cz _2$ identification in the phase space
reads; $(\m ,\l ,a,b)\sim -(\m ,\l ,a,b)$.
 The Hamiltonian is identically zero. Since all local
symmetries have been taken care of, the only remaining
gauge redundancies are the large diffeomorphisms and the large $O(1,2)$
transformations. The large diffeomorhisms on the torus form a
group -- the modular group, $PSL(2,\cz )=SL(2,\cz)/\{{\bf 1},{\bf -1}\}$ --
which is generated by
\bea S: && (a,\l )\rightarrow (b,\m ),\hspace{10mm} (b,\m )\rightarrow
(-a,-\l ) \label{S}\\
T:&& (a,\l )\rightarrow (a,\l ), \hspace{10mm} (b,\m ) \rightarrow (b+a,\m +
\l ) \label{T}\eea

while the remaining large $O(1,2)$ part is isomorphic to $\cz _2$ and
generated by
\be O:\hspace{5mm}(a,\l )\rightarrow (a,-\l ),\hspace{10mm} (b,\m
)\rightarrow (b,-\m ) \label{otransf} \ee
To quantize this system, I will follow the algebraic quantization program
outlined in \cite{tate} while treating the above large symmetries as gauge.
Alternatively, one may see the quantization below as a straightforward
application of standard geometrical quantization \cite{woodhouse} with a
special polarization of the phase space.
Instead of quantizing the fundamental Poisson-algebra in the vector
space of complex valued functions in $\cq $, I choose to quantize the
Poisson-algebra of the phase space functions\footnote{The reasons for this
choice is partly that these functions are good coordinates on phase space --
the Jacobian is everywhere non zero -- and partly that the modular group now
has a nice action in $m,\bar{m}$ space except for the region $m-\bar{m}=0$
$\Leftrightarrow $ $a\m - b\l =0$.}
\bea m:=\frac{(b+ i\m )}{(a+ i \l )}&& \bar{m}:=\frac{(b-i \m )}{(a- i \l )}
\label{m}\\
p:=-i (a- i \l )^2 && \bar{p}:=i (a+ i \l )^2 \label{p}\eea
in the vector space of complex valued functions of $m$ and $\bar{m}$. Note
that these functions automatically takes care of the redundancy
identification $(\m , \l ,a,b)\sim -(\m , \l ,a,b)$, while the large
diffeomorphisms and $O(1,2)$ transformations become
\bea S:&& m\rightarrow -\frac{1}{m},\hspace{10mm} p\rightarrow m^2p
\label{S2}\\
T: && m\rightarrow m+1,\hspace{10mm} p\rightarrow p \label{T2} \\
O: && m\rightarrow \bar{m},\hspace{10mm} p\rightarrow -\bar{p}\label{O2} \eea
and similarly for their complex conjugate functions.
Explicitly, the classical Poisson-algebra is
\bea && \{m,\bar{m}\}=\{m,p\}=\{\bar{m},\bar{p}\}=\{p,\bar{p}\}=0 \\
&& \{m,\bar{p}\}=\{\bar{m},p\}=-2 \eea
which is easily quantized by the representation
\be \hat{m}=m,\hspace{5mm}\hat{\bar{m}}=\bar{m},\hspace{5mm}
\hat{p}=2i\frac{\partial}{\partial \bar{m}} + \frac{1}{m_2}, \hspace{5mm}
\hat{\bar{p}}:=2i
\frac{\partial}{\partial m} - \frac{1}{m_2} \ee
where $m=m_1 + im_2$. (The representation of $\hat{p}$ and $\hat{\bar{p}}$ is
chosen such that formally $\hat{p}^\dagger = \hat{\bar{p}}$ w.r.t the
inner-product
defined below.)

Next we need to find the physical states, {\it i.e.} complex valued functions
of $m$ and $\bar{m}$ that are invariant under the remaining "gauge
transformations" (\ref{S2}), (\ref{T2}) and (\ref{O2}). Since we know that
the quotient space ${\bf \cc} /PSL(2,\cz )/\cz _2 $ simply is the torus moduli
space\footnote{Actually, to be completely honest, this is only true
for the complex plane minus the real axis. However, since the real axis is a
region of measure zero in the complex plane, I will simply neglect it.},
$\cm $
 \cite{riemm}, we automatically know that our physical states are given by
functions on this space. Then, we need an inner-product, and since there
exist a unique -- up to scaling -- modular invariant measure \cite{riemm} on
the moduli space, the inner product is naturally chosen to be

\be <\J _1| \J _2 >:=\int _{\cm } \frac{dm_1 dm_2}{m_2^2}\bar{\J }_1\J _2
 \ee
Explicitly, the moduli space is $\cm $: $|m|\geq 1$, $m_2 >0$ and
$|m_1|\leq \frac{1}{2}$.

To complete the quantization one should now find
a complete set of $PSL(2,\cz )\otimes \cz _2$ invariant physical observables,
check that they are
self adjoint in the Hilbert space, and check if their algebra is irreducibly
represented in the Hilbert space. I will not address these issues here.

Thus, this completes the construction of a non-trivial Hilbert space carrying
a one-dimensional unitary representation of the modular group.\\ \\
Now, how is this quantum theory related to the ADM-quantization? Just by
comparing the definitions (\ref{m})-(\ref{p}) to the time dependent canonical
transformation between half the phase space\footnote{Explicitly we have $\G
_{ADM} \cong \G _{conn}/\cz _2$, where $\G $ denote the phase space and $\cz
_2$ acts in $\G _{conn}$ as in eq. (\ref{otransf}).} of the connection
formulation and
the ADM-formulation \cite{carlip1}, one immediately sees that the functions
$m$, $\bar{m}$, $p$ and $\bar{p}$ are nothing more than the ADM-variables at
a specific instant of ADM-time. Thus, the above
quantum theory is simply the Heisenberg picture of the ADM
Schr\"{o}dinger picture, and unitary equivalence follows directly.\\ \\
Finally we may note that although we have managed to construct a non trivial
modular invariant theory starting from the connection formulation, our
quantum theory is not a connection representation \cite{ashtekar1, marolf}.
This immediately implies that we do not have any obvious way to relate this
quantum theory to the so called loop-representation \cite{ashtekar1, marolf}.
 Furthermore,
it seems very unlikely that the above construction should have any chance of
working in the time-like sector where the modular group does not act properly
discontinuous even in the full phase space. Thus the construction of a non
trivial
modular invariant connection representation still remains as an open
problem.\\ \\
{\bf Acknowledgements}\\
I thank Abhay Ashtekar for suggesting solving the problem of ref.\cite{pp} by
choosing a different polarization. I also thank Don Marolf for discussions.

\end{document}